\documentclass[preprint2]{aastex}

\usepackage{color}                       
\usepackage{url}                         
\usepackage{hyperref}
\hypersetup{colorlinks, citecolor=blue, filecolor=blue, linkcolor=blue, urlcolor=blue}
\usepackage{epstopdf}

\usepackage{bm}
\makeatletter

\newcommand{\Rmnum}[1]{\expandafter\@slowromancap\romannumeral #1@}
\newcommand{\rsun}{R$_{\odot}$}
\makeatother

\newcommand{\corr}[1]{{{#1}}}

\shorttitle{Shock wave energetics}
\shortauthors{D.M.~Long et al.}

\begin{document}

\title{The energetics of a global shock wave in the low solar corona}

\author{David M. Long, Deborah Baker, David R. Williams}
\affil{Mullard Space Science Laboratory (UCL), Holmbury St. Mary, Dorking, Surrey, RH5 6NT, UK}
\author{Eoin P. Carley, Peter T. Gallagher, Pietro Zucca}
\affil{Astrophysics Research Group, School of Physics, Trinity College Dublin, Dublin 2, Ireland}
\email{david.long@ucl.ac.uk}

\begin{abstract}
As the most energetic eruptions in the solar system, coronal mass ejections (CMEs) can produce shock waves at both their front and \corr{flanks} as they erupt from the Sun into the heliosphere. However, the amount of energy produced in these eruptions, and the proportion of their energy required to produce the \corr{waves}, is not well characterised. Here we use observations of a solar eruption from 2014~February~25 to estimate the energy budget of an erupting CME and the globally--propagating ``EIT wave'' produced by the rapid expansion of the CME flanks in the low solar corona. The ``EIT wave'' is shown using a combination of radio spectra and extreme ultraviolet images to be a shock front with a Mach number greater than one. Its initial energy is then calculated using the Sedov--Taylor blast--wave approximation, which provides \corr{an approximation} for a shock front propagating through a region of variable density. This approach provides an initial energy estimate of $\approx$2.8$\times$10$^{31}$~ergs to produce the ``EIT wave'', which is \corr{approximately 10~\% the kinetic energy of the associated CME (shown to be $\approx$2.5$\times$10$^{32}$~ergs)}. These results indicate that the energy of the ``EIT wave'' \corr{may be} significant and must be considered when estimating the total energy budget of \corr{solar eruptions}.
\end{abstract}


\section{Introduction}
\label{sect:intro}

Typically observed as globally--propagating, diffuse features which can traverse the solar disk in under an hour, ``EIT waves'' were first observed in the solar corona using the Extreme ultraviolet Imaging Telescope \citep[EIT;][]{Delaboudiniere:1995a} onboard the \emph{Solar and Heliospheric Observatory} \citep[\emph{SOHO};][]{Domingo:1995a} spacecraft. These initial observations \citep{Moses:1997a,Dere:1997a,Thompson:1998a} were quickly followed by many more, first from \emph{SOHO}/EIT \citep[e.g.,][]{Thompson:2009a}, then subsequently from the \emph{Solar Terrestrial Relations Observatory} \corr{\citep[\emph{STEREO}; e.g.,][]{Long:2008a,Veronig:2008a,Long:2011a}} and more recently the \emph{Solar Dynamics Observatory} \corr{\citep[\emph{SDO}; e.g.,][]{Liu:2010a,Long:2011b}} spacecraft. Although originally considered to be the coronal counterpart of the chromospheric Moreton--Ramsey wave \citep{Moreton:1960a,Moreton:1960b} as predicted by \citet{Uchida:1968a}, discrepancies in their kinematics and morphology \corr{raised difficulties with} this assumption \citep[e.g.,][]{Delannee:2000a}. 

Traditionally, ``EIT waves'' have been interpreted either as waves \corr{\citep[fast--mode magnetoacoustic waves, shock waves or, alternatively, soliton waves e.g.,][]{Thompson:2000a,Wang:2000a,Wills-Davey:2007a,Grechnev:2008a}} or as a brightening resulting from the restructuring of the coronal magnetic field \corr{\citep[e.g.,][]{Delannee:2000a,Chen:2002a,Attrill:2007a,Delannee:2008a}} during the eruption of a coronal mass ejection (CME). More recently, they have been interpreted using a hybrid approach, where the observed ``EIT wave'' is a freely--propagating magnetoacoustic wave initially driven by the rapid expansion of the CME as it erupts into the heliosphere \citep[cf.,][]{Zhukov:2004a,Downs:2011a}. A more detailed discussion of the different theories proposed to explain ``EIT waves'' and the observations leading to these interpretations may be found in the recent reviews by \corr{\citet{Wills-Davey:2009a,Gallagher:2011a,Zhukov:2011a}} and \citet{Liu:2014a}.

\begin{figure*}[!ht]
\centering
\includegraphics[width = 0.99\textwidth, trim=10 5 0 35,clip=]{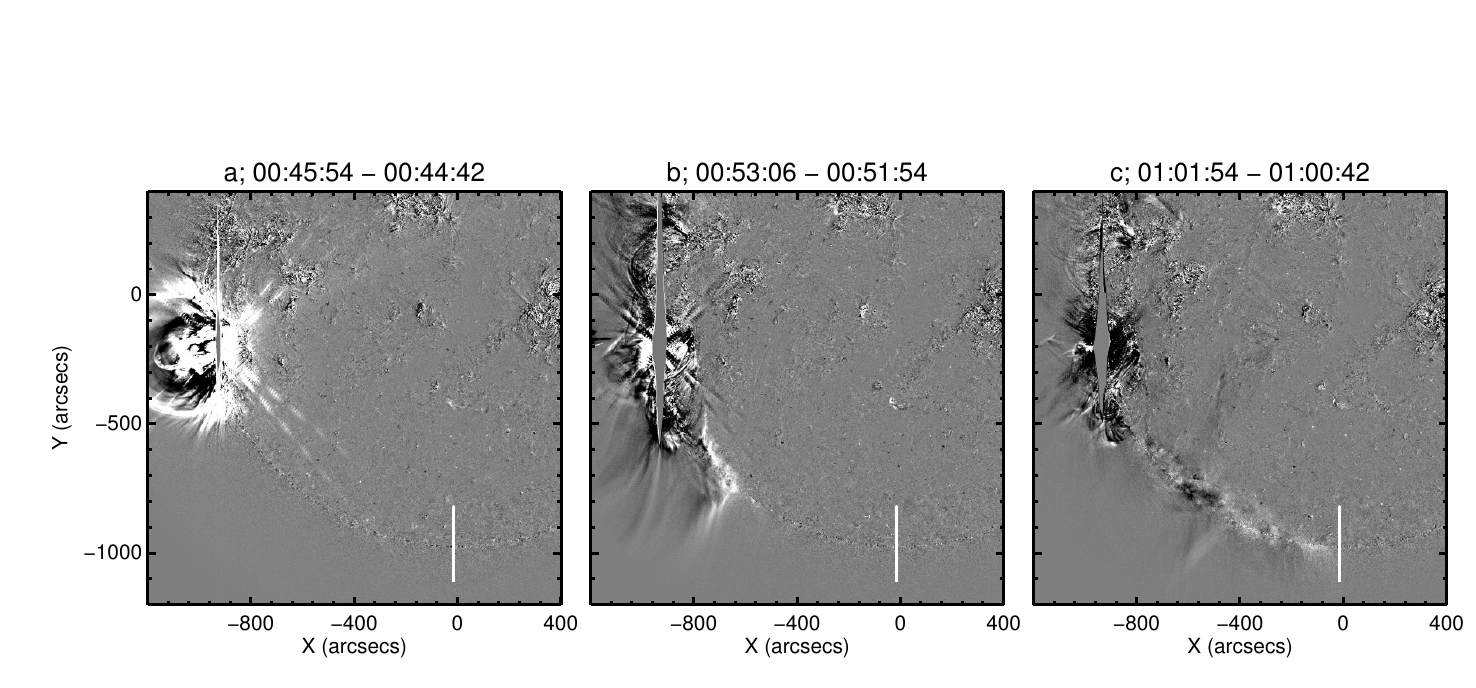}
\caption{Running-difference images from the 193~\AA\ passband showing the propagation of the global wave--pulse at 00:45:54~UT (panel~a), 00:53:06~UT (panel~b) and 01:01:54~UT (panel~c). The EIS slit located at the southern polar coronal hole is indicated in each panel.}
\label{fig:wave_prop}
\end{figure*}

Although primarily observed as broad, diffuse features, there have been observations of sharply defined ``EIT waves'' which are usually identified as shock waves \corr{\citep[e.g.,][]{Thompson:2000a,Grechnev:2008a,Grechnev:2011a,Veronig:2010a,Ma:2011a}}. These ``S-waves'' \citep[e.g.,][]{Biesecker:2002a} have been noted to exhibit a drop in intensity\corr{, as well as a dramatic change in morphology with propagation} \citep{Warmuth:2004a} and may be evidence for a shock wave decaying to an ordinary magnetohydrodynamic wave as it propagates away from its source \citep[cf.][]{Warmuth:2007a}. This is consistent with recent observations by \citet{Carley:2013a}, which showed evidence of acceleration of particles by a shock wave, driven in turn by an erupting CME. The shock wave was observed in the low corona as an ``EIT wave'' using EUV data from the Atmospheric Imaging Assembly \citep[AIA;][]{Lemen:2012a} onboard \emph{SDO}. The paucity of shock wave observations compared to those of the more general ``EIT wave'' observations is consistent with the classification system proposed by \citet{Warmuth:2011a}, which groups ``EIT waves'' based on distinct kinematic classes. These include initially fast waves with pronounced deceleration, waves with moderate and nearly constant speeds and slow disturbances exhibiting erratic kinematics. 

The very high temporal and spatial resolution of \emph{SDO}/AIA has greatly increased the number of observations of ``EIT waves'' \citep[e.g.,][]{Nitta:2013a}, and has contributed to a deeper understanding of the phenomenon \citep[as evidenced by the detailed recent review by][]{Liu:2014a}. However a number of questions remain, particularly with regard to the energy budget of the wave and how this is related to the energy of the associated flare and CME. Previous work, particularly by \citet{Emslie:2004a,Emslie:2012a}, has attempted to examine the partition of energy in a solar eruption, finding that the CME dominates the energy budget with an associated kinetic energy of $\approx$10$^{32}$~ergs for each of the events studied. However, neither of these discussions included the energy of an ``EIT wave''. 

Although not included in analysis of the fractionation of energy budgets, some work has been done on identifying the energy associated with the ``EIT wave'' itself. A lower bound on the energy of an ``EIT wave'' was proposed by \citet{Ballai:2005a} using the oscillation profile of a coronal loop which had been impacted by a propagating ``EIT wave''; this was estimated to be $\approx$3.4$\times10^{18}$~J (3.4$\times10^{25}$~ergs). A more recent analysis of the conductive, kinetic and thermal energy of the ``EIT wave'' was performed by \citet{Patsourakos:2012a}, whose \corr{generic} estimate of the energy of the ``EIT wave'' \corr{using typical observational parameters} is much higher, at $\approx$1.8$\times10^{29}$~ergs, comparable to the energy of a small flare. The relationship between the energy of the propagating wavefront and the energy of the CME itself remains uncertain, although \citet{Liu:2012a} did use observations of a filament--cavity oscillation resulting from the impact of an ``EIT wave'' to estimate the energy of the wavefront at $\approx$10$^{28}$--10$^{29}$~ergs, much smaller than the kinetic energy of the associated CME ($\approx$10$^{31}$~ergs). This estimate of an ``EIT wave'' having $\approx$1~\% the energy of the associated CME is much lower than the 10~\% proposed by \citet{Vourlidas:2010a}, suggesting that more work is required on this topic.

In this paper, we present observations of an ``EIT wave'' propagating through the solar corona made by \emph{SDO}/AIA. Section~\ref{sect:obs} presents the observations of the ``EIT wave'', which are shown in Section~\ref{sect:shock} to be consistent with that of a shock wave initially driven by the CME. This allows the energy budget of the shock to be calculated in Section~\ref{sect:energy} before the implications of these observations are discussed in Section~\ref{sect:disc}.

\section{Observations and Data Analysis}
\label{sect:obs}

The ``EIT wave'' studied here erupted from NOAA active region AR~11990 close to the East limb of the Sun on 2014~February~25 (as shown in Figure~\ref{fig:wave_prop}) and was associated with a GOES X4.9 class flare which began at 00:39~UT before peaking at 00:49~UT. The eruption was also associated with a CME and a Type~\Rmnum{2} radio burst, indicating the presence of a shock. Although the ``EIT wave'' pulse may be seen in all of the passbands observed by \emph{SDO}/AIA, the 193~\AA\ passband was primarily used in this analysis as it provided the clearest observations of the wave--pulse. \corr{Despite the intense nature of the erupting flare and the broad extent of the associated CME, the wave--pulse did not propagate equally in all directions across the Sun, being constrained by the locations of the neighbouring active regions. However, it was clearly observed to propagate south from the erupting active region through the quiet Sun towards the southern polar coronal hole.}

The AIA data were prepared using the normal \textsf{aia\_prep.pro} routines available from SolarSoft \citep{Freeland:1998a}. The intense nature of the erupting X4.9 flare meant that the data were strongly affected by the Automatic Exposure Control (AEC) algorithm which dynamically changes the exposure time of the images depending on the intensity of the most recent image. This can produce a constant-exposure time-series with a number of low-exposure-time images interleaved. As faint features in particular are difficult to discern in images with a low signal-to-noise, images with a lower than normal exposure time were not considered in this analysis. 

At the time of the eruption, the Extreme ultraviolet Imaging Spectrometer \citep[EIS;][]{Culhane:2007a} onboard the \emph{Hinode} \citep{Kosugi:2007a} spacecraft was observing the southern polar coronal hole using a study with short exposure times ($\sim$30~s), which reduced the signal-to-noise of the observations. Despite this, it was possible to use the \ion{Fe}{12} 195.12~\AA\ emission line (which is the strongest line observed by \emph{Hinode}/EIS) to examine the observed ``EIT wave''. This line is also observed by the 193~\AA\ passband on \emph{SDO}/AIA, allowing a comparison between the two instruments.

\section{Characterisation of the wave--pulse}
\label{sect:shock}

\begin{figure}[!t]
\centering
\includegraphics[width = 0.48\textwidth, trim=0 0 0 0,clip=]{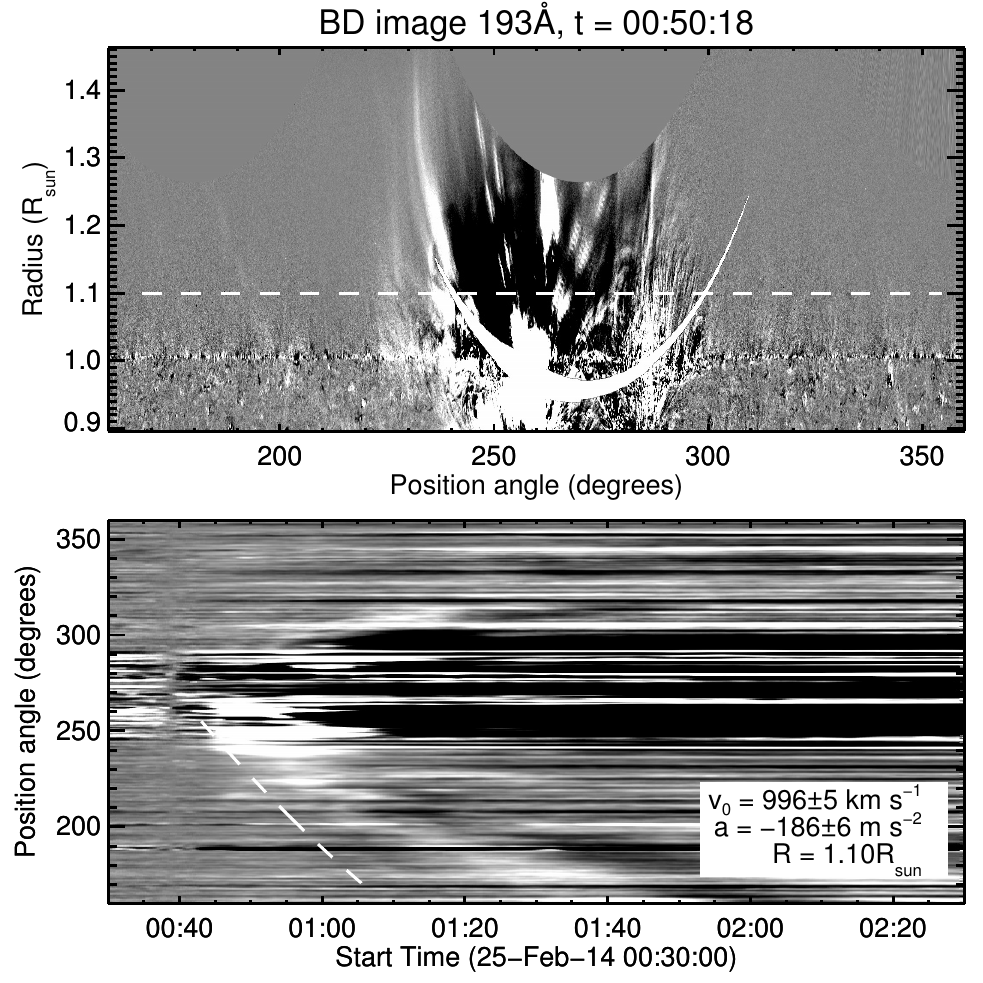}
\caption{\emph{Top}: Base-difference deprojected annulus image showing the eruption at 00:50:18~UT in the 193~\AA\ passband. The vertical axis shows the height from Sun-centre in solar radii while the horizontal axis shows the angle clockwise from solar north. \emph{Bottom}: Base-difference image showing the temporal variation at a height of 1.1~\rsun\ (indicated by the dashed white line in the upper panel). The dashed white line shows the fit to the shock front \corr{at the leading edge of the bright feature}, with the fitted initial velocity and acceleration given in the bottom right of the panel.}
\label{fig:polar_plots}
\end{figure}

Considering the many interpretations for ``EIT waves'' and the different physics associated with each theory, it was important to characterise the true physical nature of the wave--pulse prior to any further analysis. \corr{The variations in Alfv\'{e}n speed within the} neighbouring active regions meant that the wave--pulse could not easily propagate through them and instead was forced to propagate primarily north and south along the solar limb as shown in Figure~\ref{fig:wave_prop} and the associated online movie movie1.mov. \corr{As a result, it was not} possible to use the CorPITA technique \citep{Long:2014a} to identify and characterise the wave--pulse. Instead, a deprojected annulus of $\approx$0.5~\rsun\ width was used to study the evolution of the wave--pulse along the solar limb. This may be seen in the top panel of Figure~\ref{fig:polar_plots} which shows the deprojected annulus image at 00:50:18~UT. Note that this is a base--difference image, with a pre--event image at 00:39:00~UT subtracted in order to highlight the wave--pulse. A similarly base--differenced stack plot is shown in the bottom panel of Figure~\ref{fig:polar_plots}, which gives the variation in the lateral extent of the wave--pulse with time at a height of 1.1~\rsun\ (indicated by the dashed line in the top panel).

\begin{figure}[!t]
\centering
\includegraphics[width = 0.485\textwidth, trim=10 5 0 0,clip=]{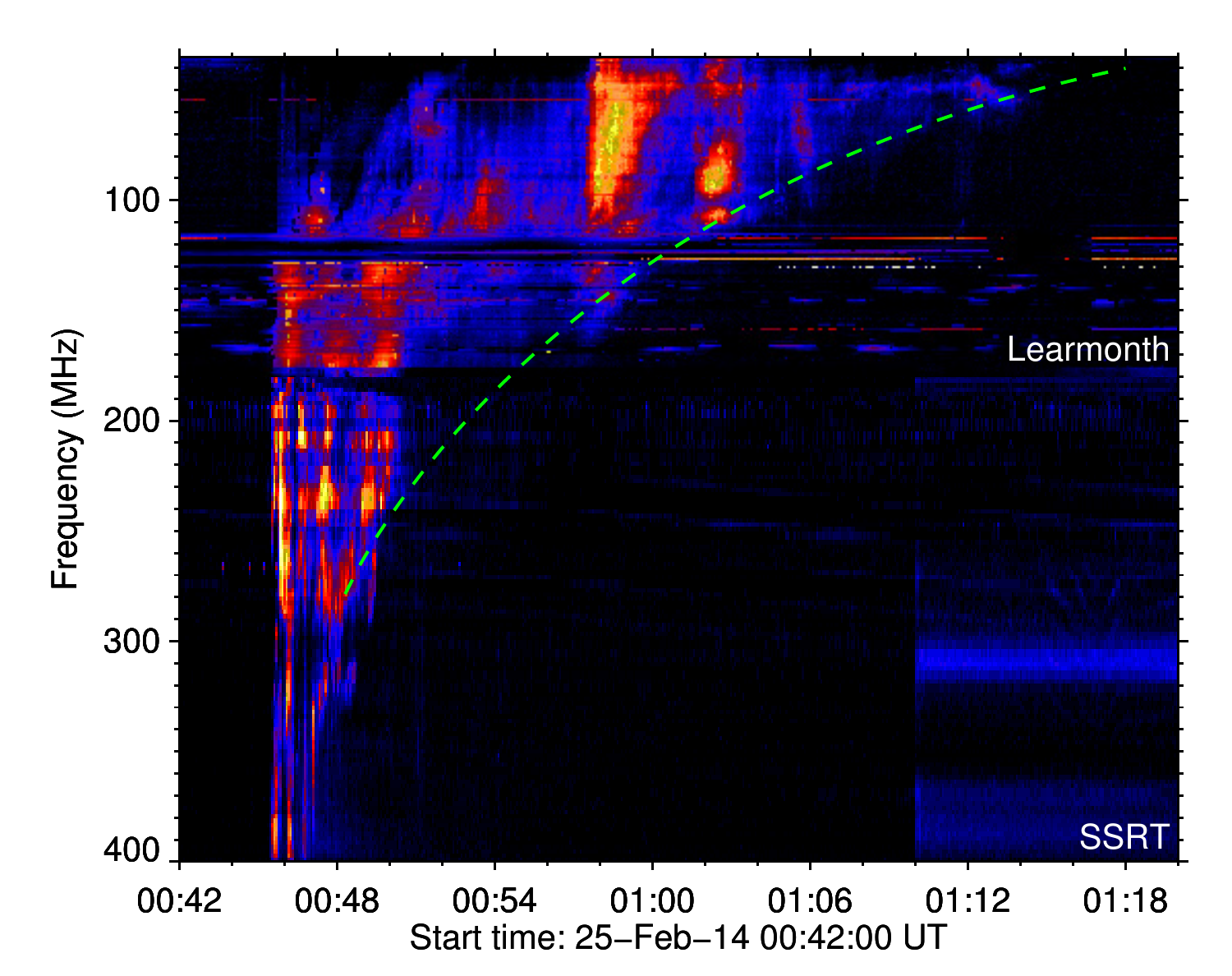}
\caption{Radio spectra from the \corr{Learmonth Solar Radio Spectrograph (25--180~MHz) and the} Siberian Solar Radio Telescope \corr{(SSRT; 180--400~MHz)} showing the Type~\Rmnum{2} radio burst associated with this event. \corr{The Type~\Rmnum{2} burst is indicated by the dashed line.}}
\label{fig:radio}
\end{figure}

The EUV observations of the wave--pulse as shown in Figure~\ref{fig:wave_prop} show a clearly defined bubble with a sharp leading edge associated with the CME eruption. Although these observations are consistent with the interpretation of the wave--pulse as a shock wave, it was necessary to examine the kinematics of the wave--pulse to test this assumption. This was done by examining the variation in intensity over time, at a fixed height of 1.1~\rsun, as shown in the stack plot in Figure~\ref{fig:polar_plots}. The leading edge of the bright feature was identified and fitted using a quadratic model (dashed line in the bottom panel of Figure~\ref{fig:polar_plots}), with this process repeated 10 times to minimise uncertainty. The dashed white line here indicates the fit to the leading edge of the shock front and has an estimated initial velocity of $v_{0}=996\pm5$~km~s$^{-1}$ and an estimated acceleration of $a=-186\pm6$~m~s$^{-2}$. 

This approach was used to find the kinematics of the wave--pulse across a range of heights from 1~\rsun\ to 1.3~\rsun, giving an initial velocity range of $681<v_{0}<1233$~km~s$^{-1}$ with a mean of $v_{0}=920\pm119$~km~s$^{-1}$, and an acceleration range of $-479<a<97$~m~s$^{-2}$ with a mean of $a=-169\pm156$~m~s$^{-2}$. These fitted initial velocities are larger than the $\sim$200--400~km~s$^{-1}$ range estimated by \citet{Thompson:2009a} \corr{using observations from \emph{SOHO}/EIT} and the mean velocity of $\sim$644~km~s$^{-1}$ measured by \citet{Nitta:2013a} \corr{using higher--cadence observations made by \emph{SDO}/AIA}, indicating that this was a particularly fast ``EIT wave'' event.

\begin{figure*}[!htp]
\includegraphics[width = 1\textwidth,trim=0 0 0 0,clip=]{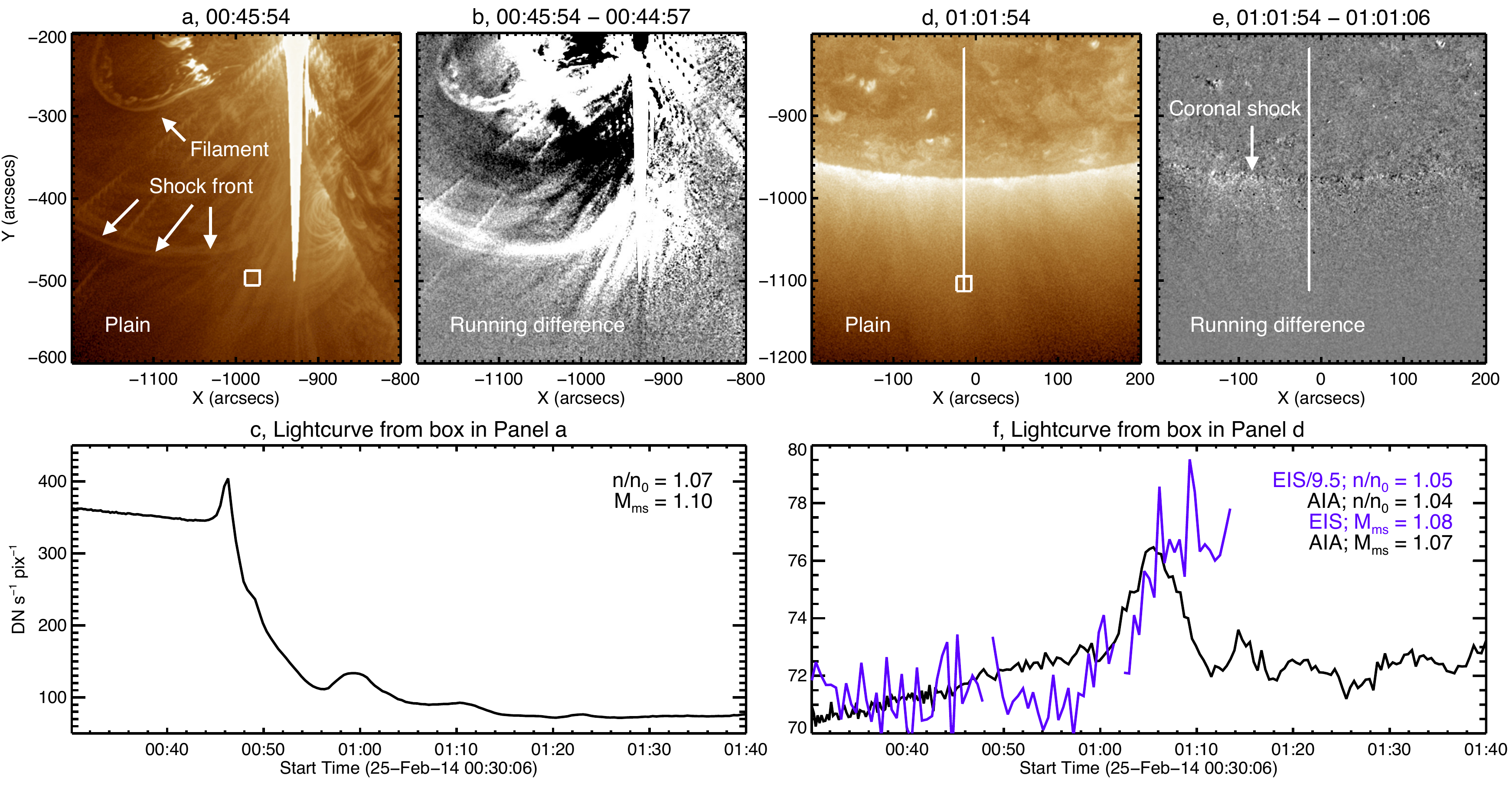}
\caption{The eruption from 2014~February~25 as observed by the \corr{\emph{SDO}/AIA} 193~\AA\ passband. The eruption is shown close to the source using \corr{intensity} and running-difference images in panels~a \& b respectively and over the South Pole using \corr{intensity} and running-difference images in panels~d \& e respectively. Panels~c \& f show the intensity profiles from the white square regions shown in panels~a \& d. The \emph{Hinode}/EIS field-of-view is shown in panels~d \& e, with the intensity profile from the EIS pixels within the white box shown in blue in panel~f. The \corr{intensity} images in panels~a \& d were processed for display here using the Multi-scale Gaussian Normalisation (MGN) technique of \citet{Morgan:2014a}.}
\label{fig:erupt_loc}
\end{figure*}

In addition to the EUV observations, a Type~\Rmnum{2} radio burst associated with this event was observed by \corr{both the Learmonth Solar Radio Spectrograph \citep{Lobzin:2010a} and} the Siberian Solar Radio Telescope \citep{Smolkov:1986a} in the period 00:46--\corr{01:06} UT (see Figure~\ref{fig:radio}). Although quite complex and fragmented, the burst shows clear characteristics of a Type~\Rmnum{2}, with a slow frequency drift from $\sim$400 to $\sim$100~MHz. The height of the shock was estimated by assuming that we observe plasma emission and can therefore convert directly to electron density using \corr{$f_{p} = 8980\sqrt{n_e}$}. The radial stratification of density with height was then estimated using density measurements derived from all of the EUV filters of AIA as described by \citet{Zucca:2014a}. This approach allows the height that the radio emission was produced at to be inferred from its frequency. In this case, the Type~\Rmnum{2} burst starting frequency of 350~MHz placed it at an initial height of 1.1~\rsun (assuming the emission is the 2nd harmonic of the plasma frequency). Furthermore, a velocity of $\approx$550~km~s$^{-1}$ could be derived from its frequency drift assuming radial propagation. This identification of a shock at low heights in the corona is consistent with the assertion of the ``EIT wave'' being a shock. However, while it is possible that the ``EIT wave'' and Type~\Rmnum{2} burst are from the same MHD disturbance in the corona, they may be at two spatially distinct locations and propagating in different directions. We are therefore cautious not to compare the Type~\Rmnum{2} and ``EIT wave'' velocities as they are not generally expected to be the same \citep{Klassen:2000a}, instead using the presence of a Type~\Rmnum{2} burst as a positive identification of a shock in the low corona.


The high measured velocities of the ``EIT wave'' and the occurrence of a Type~\Rmnum{2} burst are all consistent with the presence of a shock front, indicating that it may be possible to estimate the magnetosonic Mach number M$_{ms}$ of the disturbance. This can be achieved using the Rankine--Hugoniot relations for a perpendicular magnetosonic shock ($\theta=90^{\circ}$), described as,
\begin{equation}
M_{ms} = \sqrt{\frac{X(X + 5 + 5\beta)}{2(4 - X)}}, \label{eqn:mach}
\end{equation}
where $X$ is the density compression ratio $n/n_0$, the plasma-$\beta=0.1$ and an adiabatic index $\gamma=5/3$ has been assumed \citep{Priest:1982a,Vrsnak:2002a}. The density compression ratio $n/n_0$ can be estimated from EUV images by examining the variation in image intensity as the shock front passes through the field of view as,
\begin{equation}
\frac{n}{n_0} = \sqrt{\frac{I}{I_0}}, \label{eqn:dens}
\end{equation}
where $I_0$ and $n_0$ are the pre--event image intensity and density respectively while $I$ and $n$ are the image intensity and density measured during the passage of the wave--pulse. This relationship was used to calculate the variation in image intensity close to the source of the eruption and also at the southern polar coronal hole as shown in Figure~\ref{fig:erupt_loc}. This approach has previously been employed by \citet{Muhr:2011a} and \citet{Zhukov:2011a} and allows an estimate to be made of Mach number assuming no strong variation in temperature and a constant density along the depth of emission (assumed to be the pressure scale height). Since the total emission will be due to the emission measure along the line of sight, only a fraction of which is occupied by the plasma experiencing the shock, Equation~\ref{eqn:dens} is therefore an underestimate of the true density enhancement.

A clear jump in intensity is apparent in panel~c of Figure~\ref{fig:erupt_loc}, which gives the intensity profile for the white box shown in \corr{panels~a and~b}. This enhancement in image intensity is mirrored in the corresponding density compression ratio of $n/n_0=$1.07 as given in the legend. The resulting magnetosonic Mach number of $M_{ms}=$1.10 confirms that the observed ``EIT wave'' was shocked at this stage in its propagation. 

This process was repeated close to the southern polar hole as shown in panels~d, e \& f of Figure~\ref{fig:erupt_loc}, with the intensity variation measured in a location at the bottom of the \emph{Hinode}/EIS slit as shown in panel~d. The temporal variation in the measured intensity was much lower here, likely reflecting the distance from the source. The density compression ratio of $n/n_0=$1.04 and the resulting magnetosonic Mach number of $M_{ms}=$1.07 for the 193~\AA\ passband are lower than close to the source but still exceed unity, indicating that while the wave--pulse remains shocked, the shock has decayed with propagation \citep[cf.][and references therein]{Kantrowitz:1966a,Mann:1995a,Warmuth:2007a}.

The position of the EIS slit allowed the intensity ratio and hence density compression ratio and Mach number to be compared between the different instruments. The intensity of the EIS 195.12~\AA\ emission line was averaged along the EIS pixels corresponding to the white box shown in Figure~\ref{fig:erupt_loc}, producing the blue intensity profile shown in panel~f of Figure~\ref{fig:erupt_loc} (note that the EIS intensity has been scaled to match the AIA intensity in the plot). Both intensity profiles exhibit similar variations, indicating that the intensity increase is a true feature. Although the EIS intensity exhibits more random variation with time compared to AIA, this is most likely due to the lower signal-to-noise in the narrow-band (monochromatic) EIS signal as opposed to the higher signal-to-noise in the broadband AIA signal. The density compression ratio and resulting Mach number were then calculated for the EIS intensity profile using Equations~\ref{eqn:mach} and \ref{eqn:dens} and were found to be 1.05 and 1.08 respectively. These are comparable to those measured with AIA and provide independent verification of both the approach and the results. 

\begin{figure}[!t]
\centering
\includegraphics[width = 0.5\textwidth, trim=10 5 0 0,clip=]{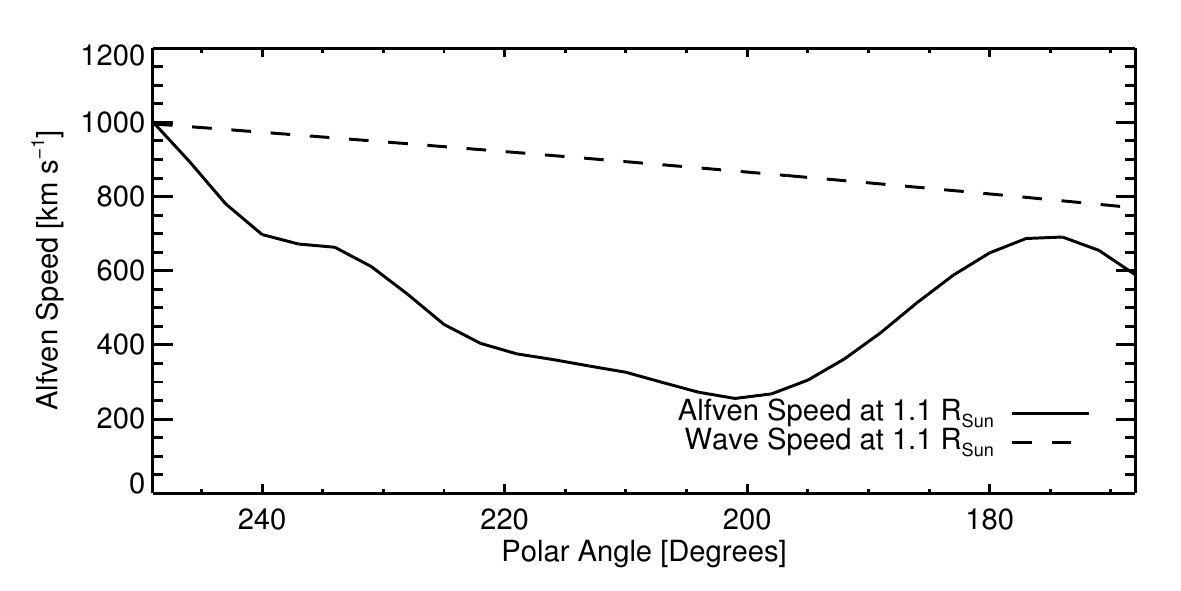}
\caption{The Alfv\'{e}n speed at a height of 1.1~\rsun\ as derived using the technique of \citet{Zucca:2014a}. The fitted kinematics of the shock front are indicated by the dashed line.}
\label{fig:alfven_prof}
\end{figure}

These numbers may be further examined by comparing the velocity of the wave--pulse to the local Alfv\'{e}n speed, which can be taken as a proxy measure of the fast-mode speed. As described by \citet{Zucca:2014a}, the local Alfv\'{e}n speed may be calculated using a combination of emission measure and polarisation brightness. The local Alfv\'{e}n speed of the plasma that the wave--pulse propagated through (at a height of 1.1~\rsun) is shown in Figure~\ref{fig:alfven_prof} to decrease from an initial peak of $\approx$900~km~s$^{-1}$ near the source to under $\approx$300~km~s in the quiet corona at a position angle of $\sim$200$^{\circ}$ clockwise from solar north. It then increases to $\approx$600~km~s$^{-1}$ at the southern coronal hole. The speed of the wave--pulse (indicated by the dashed line in Figure~\ref{fig:alfven_prof}) exceeds the local Alfv\'{e}n speed at all locations, indicating that the wave--pulse was a shock. However, the resulting Mach number where the intensity ratio was measured near the source and close to the polar coronal hole was quite low, implying that it was very weakly shocked in these locations and that the assumption of negligible temperature variation is valid in this case.

\section{Energetics of the Eruption}
\label{sect:energy}

\begin{figure}[!t]
\centering
\includegraphics[width = 0.49\textwidth,trim=15 5 0 0,clip=]{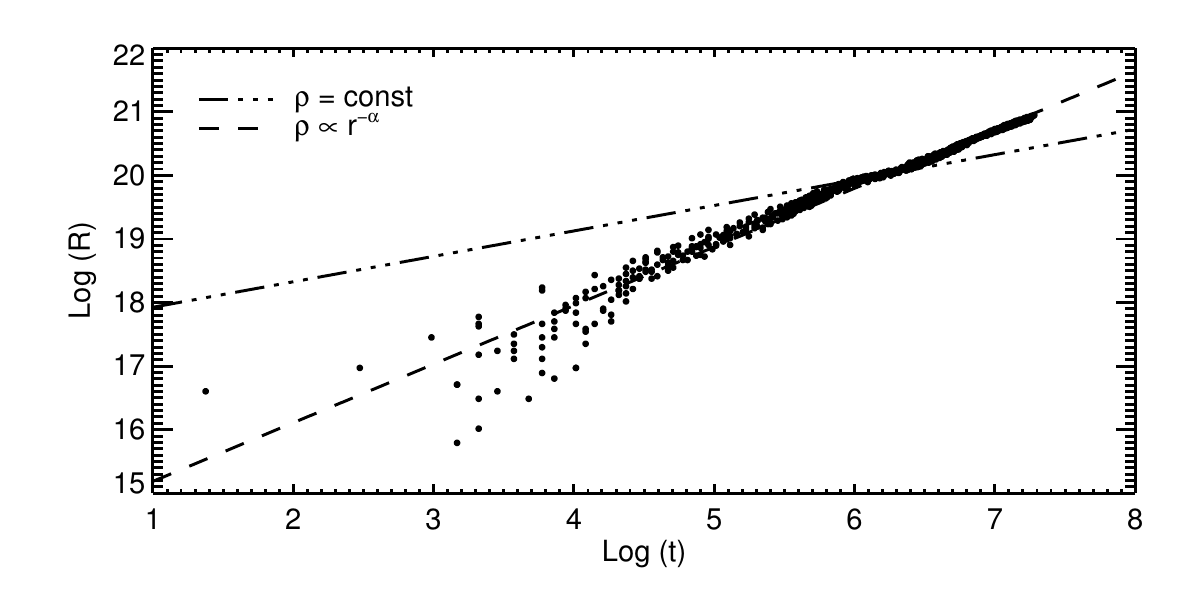}
\caption{Plot of log~$R$ versus log~$t$ for the distance of the shock front from the source at a height of 1.1~\rsun. The constant density fit is shown by the dot--dash line with the dashed line indicating the fit assuming a variable-density medium ($\rho \propto r^{-\alpha}$ with $\alpha=2.78$).}
\label{fig:energy}
\end{figure}

With the observed wave--pulse identified as a propagating shock wave, it was possible to estimate its energy using the Sedov--Taylor relation \citep{Sedov:1946a,Taylor:1950a}. This technique was originally developed during the Second World War to estimate the energy of an atomic bomb, and assumes a spherical blast wave emanating from a point source. \corr{It should be noted at this point that this interpretation may not be strictly correct in this case as the shock wave may be initially driven by the lateral expansion of the associated CME rather than originating from a point source such as a flare \citep[cf.][]{Patsourakos:2010a}. However, it serves as a good first--order approximation, allowing an initial estimate of the energy associated with the shock front. The Sedov--Taylor relation} may be derived using dimensional analysis and, assuming that the shock front is passing through a region of constant density, can be written in terms of the radius $R$ at time $t$ as,
\begin{equation}
R(t) \approx \left(\frac{E t^{2}}{n}\right)^{1/5},
\end{equation}
where $E$ is the energy of the blast wave and $n$ is the density. This may then be used to obtain an estimate of the energy E by plotting log~$R$ against log~$t$ and fitting the result with a line of slope $2/5$ \citep[cf.][]{Taylor:1950b}. This is shown in Figure~\ref{fig:energy}, which shows log~$R$ plotted against log~$t$ at a heliocentric distance of 1.1~\rsun. It is clear that the line of slope $2/5$ (the dashed--dotted line) is not a good fit to the data, indicating that the assumption of constant density is not valid in this case. 

An alternative approach is to consider a variable-density medium where $\rho \propto r^{-\alpha}$\corr{ (with $r$ being the distance from the source). This approach} was outlined numerically by \citet{Sedov:1959} and \corr{was} previously employed in a solar context by \citet{Grechnev:2008a}. In this case, the radius of the shock front $R$ at time $t$ is defined as,
\begin{equation}
R(t) \approx \left(\frac{E t^2}{n}\right)^{1/(5-\alpha)},
\end{equation}
where the shock decelerates if $\alpha<3$ and accelerates if $\alpha>3$. This allows $E$ to be estimated by fitting a line of the form,
\begin{equation}
\textrm{log}~R \approx \frac{2}{5-\alpha}\textrm{log}~t + \textrm{log}\left(\frac{E}{n}\right)^{1/(5-\alpha)},\label{eqn:blast}
\end{equation}
 to the data shown in Figure~\ref{fig:energy}. This provides a much better fit to the data (indicated by the dashed line in Figure~\ref{fig:energy}) for a value $\alpha=2.83$, which is consistent with the deceleration given by the fitted kinematics in Figure~\ref{fig:polar_plots}. The assumption of a variable-density medium approximates the conditions through which the shock propagates as it passes from an active region through the quiet corona towards a coronal hole. 

By assuming a typical coronal density of $n=10^8$~cm$^{-3}$, it is possible to make an estimate of the initial energy required to produce the shock front. The dashed line in Figure~\ref{fig:energy} shows Equation~\ref{eqn:blast} fitted to the data with $\textrm{log}(E/n)^{1/(5-\alpha)}\approx14.3$. This returns an energy estimate of $E\approx2.78\times 10^{31}$~ergs for this eruption, which is much larger than the \corr{energy  previously estimated using typically observed parameters} by \citet{Patsourakos:2012a} ($1.8\times10^{29}$~ergs) and \corr{the minimum energy threshold estimated by} \citet{Ballai:2005a} ($3.4\times10^{25}$~ergs). 

Although the energy value estimated here is much larger than previous estimates, it has been suggested by \citet{Vourlidas:2010a} that the energy of an ``EIT wave'' is $\approx$10~\% that of the associated CME. This suggestion may be compared with our results by estimating the energy of the associated CME using the equations
\begin{eqnarray}
E_{k} &=& \frac{1}{2}M_{cme}v^2_{cme}, \\
E_{p} &=& \frac{GM_{sun}M_{cme}}{5 R_{sun}}
\end{eqnarray}
where $E_{k}$ and E$_{p}$ are the kinetic and potential energy of the CME respectively, $G$ is the gravitational constant, $M_{sun}$ is the solar mass, $5R_{sun}$ is the height in solar radii at which the CME front first appeared in the LASCO field of view, $M_{cme}$ is the mass of the CME and $v_{cme}$ is the instantaneous velocity of the CME \citep[see e.g.,][for more details]{Carley:2012a}. These can then be combined to estimate the total mechanical energy of the CME.

The velocity of the CME was derived from a point-and-click trace of the CME front in running-difference images produced from the LASCO~C2 and C3 coronagraphs during the period 01:20--03:20~UT. There was no acceleration in the C2 and C3 fields of view (as is expected of a fast CME), and a linear fit to the heliocentric distance vs.\ time data gave a velocity of $\sim$1550~km~s$^{-1}$. The mass of the CME was derived using the Thomson scattering approach first outlined by \citet{Billings:1966a} and the methods detailed in a variety of studies such as \citet{Vourlidas:2002a,Vourlidas:2010a} and \citet{Carley:2012a}. In this case, base difference images from C2 and C3 were used and it was assumed that the CME propagated in the plane of the sky. This resulted in a mass of $\approx$2.2$\times$10$^{16}$~g, which is on the larger end of the scale of CME masses \citep{Vourlidas:2010a}. 

With the CME first appearing in the LASCO field-of-view at a height of 5~R$_{sun}$, a constant velocity and constant mass were assumed, ignoring any mass accretion by the CME. This analysis produces an estimated potential energy of $\approx$5.0$\times$10$^{30}$~ergs and kinetic energy of $\approx$2.5$\times$10$^{32}$~ergs, giving a total mechanical energy estimate for the CME of $\approx$2.6$\times$10$^{32}$~ergs. This indicates that the observed ``EIT wave'' had an energy $\sim$10~\% that of the associated CME (which had quite a large energy in terms of the overall distribution of CME energies) consistent with the results of \citet{Vourlidas:2010a}. The total mechanical energy of the eruption (excluding the energy released by the flare) can therefore be estimated at $\approx$2.8$\times$10$^{32}$~ergs.

\section{Discussion and Conclusions}
\label{sect:disc}

In this paper, we use high--cadence \emph{SDO}/AIA observations of an eruption on 2014~February~25 to estimate the energy of the associated ``EIT wave''. The ``EIT wave'' was initially produced by the erupting CME before propagating freely, although neighbouring active regions meant that it propagated primarily north and south along the limb rather than across the disk visible from Earth. This allowed the lateral motion of the wave--pulse to be used to estimate its energy using the blast wave approximation of \citet{Sedov:1946a} and \citet{Taylor:1950a,Taylor:1950b}.

The wave--pulse was identified as a shock wave using a combination of radio spectra, EUV intensity ratios, and the measured kinematics of the wave--pulse. A Type~\Rmnum{2} radio burst was observed which indicated the presence of a shock, with the high frequency of the burst suggesting a low coronal origin. The measured kinematics of the wave--pulse indicated that the observed ``EIT wave'' was particularly fast and exhibited deceleration as it propagated, suggesting that it may have propagated faster than the local fast--mode speed and was therefore a shock front. This was confirmed by examining the intensity ratio of 193~\AA\ EUV images using the technique previously discussed by \citet{Muhr:2011a} and \citet{Zhukov:2011a}. This approach allows an estimate to be made of Mach number assuming no strong variation in temperature and a constant density along the depth of emission. 

The primarily lateral motion of the wave--pulse along the limb away from the source meant that it was not possible to use the CorPITA technique of \citet{Long:2014a} to examine variations in the kinematics of the pulse. However, it was possible to use this lateral motion to estimate the initial energy budget of the wave--pulse using the blast wave approximation developed by \citet{Sedov:1946a} and \citet{Taylor:1950a,Taylor:1950b}. 

It was found that the equation of a blast wave in a region of constant density does not provide a good fit to the observed propagation of the wave--pulse. However, the equation of a blast wave in a variable-density medium provides an excellent fit to the data. This model is consistent with the observed propagation of the wave--pulse as it travels from the high density active region where it was produced through the lower density quiet corona towards the very low density coronal hole at the south pole. In addition, the degree of the variation in density was consistent with the observations of pulse deceleration, strongly indicating the validity of this approach. The resulting energy of the wave--pulse was found to be $\approx$2.8$\times$10$^{31}$~ergs, approximately 10~\% of the mechanical energy of the associated CME, which was estimated to be $\approx$2.6$\times$10$^{32}$~ergs.

\corr{This consistency between the observations and the Sedov--Taylor relation is interesting and requires further investigation. The Sedov--Taylor relation is strictly valid for a spherical blast--wave originating from a point source, which is not the case here, and as a result can be considered as providing a first--order approximation. However, a growing number of both observations and theories suggest that ``EIT waves'' are consistent with the concept of a freely--propagating shocked simple wave formed by the rapid lateral expansion of a CME in the low corona \citep[e.g.,][]{Vrsnak:2008a,Patsourakos:2010a}. In this interpretation, the timescale over which the driver acts to produce the shock wave would have to be very short to allow it to be interpreted as a blast wave. The ability of the Sedov--Taylor relation to approximate the observations here suggests that this interpretation is valid and the timescale over which the CME acts to produce the ``EIT wave'' is very short.}

\corr{This interpretation is supported by the excellent fit provided by the variable density medium formulation of the Sedov-Taylor relation. The accuracy of this model compared to the constant density formulation indicates that the wave was strongly affected by density variations in the medium that it propagated through. This would not be observed if the shock front were continuously driven by the erupting CME, suggesting that the wave front was freely--propagating.}

These results suggest that the Sedov--Taylor blast wave approximation may be used to estimate the energy of an ``EIT wave'' shock front propagating through the solar atmosphere. In addition, the energy of the ``EIT wave'' shock is comparable to the energy of a flare and is not a negligible fraction of the total energy budget \citep[cf.][]{Emslie:2004a,Emslie:2012a}. These results indicate that the energy of the ``EIT wave'' is significant and must be considered when estimating the total energy budget of a solar eruption.

\acknowledgments

\corr{The authors wish to thank the anonymous referee whose comments helped to improve the paper}. DML received funding from the European Commission's Seventh Framework Programme under the grant agreement No.~284461 (eHEROES project). DB acknowledges support by STFC Consolidated Grant ST/H00260/1. \corr{PZ is supported by a TCD Innovation Bursary. EC is supported by ELEVATE: Irish Research Council International Career Development Fellowship -- co-funded by Marie Cure Actions.} \emph{SDO}/AIA data is courtesy of NASA/\emph{SDO} and the AIA science team. \emph{Hinode} is a Japanese mission developed and launched by ISAS/JAXA, collaborating with NAOJ as a domestic partner, NASA and STFC (UK) as international partners. Scientific operation of the \emph{Hinode} mission is conducted by the \emph{Hinode} science team organized at ISAS/JAXA. This team mainly consists of scientists from institutes in the partner countries. Support for the post-launch operation is provided by JAXA and NAOJ (Japan), STFC (U.K.), NASA, ESA, and NSC (Norway).

\end{document}